\documentclass[aps,pra,twocolumn,showpacs,nofootinbib,superscriptaddress]{revtex4-1}
\usepackage{graphicx}
\usepackage{bm}
\usepackage{amssymb}
\usepackage{epstopdf}
\usepackage{amsmath}
\usepackage{amstext}
\usepackage{amsthm}
\usepackage{amsfonts}
\usepackage{latexsym}
\usepackage{hyperref}
\usepackage[margin,index]{fixme}
\usepackage[T1]{fontenc}
\usepackage{color} 

\newcommand{\bk}[2]{\langle#1|#2\rangle}

\newcommand{\ket}[1]{\left\vert#1\right\rangle}
\newcommand{\bra}[1]{\left\langle#1\right\vert}

\newcommand{\sket}[1]{|#1\rangle}
\newcommand{\sbra}[1]{\langle#1|}
\newcommand{\av}[2]{\left\langle#1\right\rangle_{#2}}

\newcommand{\idm}{\mathbb{I}}

\newcommand{\Tr}{{\rm tr}}
\renewcommand{\emph}[1]{{\it #1}}
\renewcommand{\vec}[1]{\boldsymbol{#1}}

\newcommand{\beq}{\begin{equation}}
\newcommand{\eeq}{\end{equation}}
\newcommand{\beqa}{\begin{eqnarray}}
\newcommand{\eeqa}{\end{eqnarray}}
\newcommand{\app}{\mathcal{A}}
\newcommand{\sys}{\mathcal{S}}
\newcommand{\mean}[1]{\left\langle#1\right\rangle}

\begin{document}

\title{Measuring work and heat in ultracold quantum gases}
\author{Gabriele De Chiara}
\affiliation{Centre for Theoretical Atomic, Molecular and Optical Physics Queen's University, Belfast BT7 1NN, United Kingdom}

\author{Augusto J. Roncaglia}
\affiliation{Departamento de F\'\i sica, FCEyN, UBA, Ciudad
Universitaria Pabell\'on 1, 1428 Buenos Aires, Argentina}
\affiliation{IFIBA CONICET, UBA, FCEyN, UBA, Ciudad Universitaria
Pabell\'on 1, 1428 Buenos Aires, Argentina} 

\author{Juan Pablo Paz}
\affiliation{Departamento de F\'\i sica, FCEyN, UBA, Ciudad
Universitaria Pabell\'on 1, 1428 Buenos Aires, Argentina}
\affiliation{IFIBA CONICET, UBA, FCEyN, UBA, Ciudad Universitaria
Pabell\'on 1, 1428 Buenos Aires, Argentina} 

\begin{abstract}
We propose a feasible experimental scheme to direct measure heat and work in cold atomic setups. The method is based on a recent proposal
which shows that work is a positive operator valued measure (POVM). In the present contribution, we demonstrate that the interaction between the atoms and the light polarisation of a probe laser allows us to implement such POVM. 
In this way the work done on or extracted from the atoms after a given process is encoded in the light quadrature that can be measured with a standard homodyne detection. The protocol allows one to verify fluctuation theorems and study properties of  the non-unitary dynamics of a given thermodynamic process.
\end{abstract}

\maketitle

\section{Introduction}
 The study of out-of-equilibrium thermodynamics has received a significant thrust thanks to the experimental advances in the control and manipulation of microscopic systems. From a fundamental point of view, these endeavours aim at clarifying the foundations of modern thermodynamics and its connection to information theory. From a more applied perspective, these studies aim at understanding limitations of microscopic engines and building more efficient ones. Heat and work, two ubiquitous  concepts in traditional thermodynamics, assume in this context the role of stochastic variables whose fluctuations can be ingeniously related to equilibrium properties, as is the case of the celebrated Jarzynski equality~\cite{Jarzynski}.
Many physical systems have been realized to investigate nonequilibrium thermodynamics, including  for instance strands of RNA~\cite{Liphardt}, single electron boxes~\cite{Saira}, levitated or trapped nanoparticles~\cite{Millen, Toyabe}, and colloidal particles trapped in optical potentials~\cite{Berut}.

In the last decade, general interest has been directed towards the quantum regime of out-of-equilibrium thermodynamics. In this regime, the dynamics of a small quantum system is dominated by quantum rather than thermal fluctuations. Although many open questions remain unanswered, some of the concepts of nonequilibrium classical thermodynamics have been translated into the quantum domain (see for example~\cite{CampisiRMP, Esposito}). A measure of work based on a two-measurement scheme is now commonly accepted \cite{TalknerPRE} and can be shown, for isolated systems, to fulfill a quantum extension of the Jarzynski equality \cite{Tasaki}. 
For open systems the Jarzynski equality still holds if one considers changes of energy in the system and the environment together \cite{Campisi09}. However if we consider energy changes in the system only, the fluctuation relation for the system energy ceases to work and contains a correction that depends on the properties of the environment \cite{Rastegin, Zanardi}. In fact Jarzynski equality is still valid if the corresponding evolution superoperator is {\it unital} {\cite{unital}}, i.e., if the  completely mixed state (corresponding to infinite temperatures) remains unaltered after the open system evolution.

Although implementing directly the two-measurement scheme has proven 
to be challenging, alternative routes to measure work in quantum systems 
have been proposed. One of these employs a Ramsey 
scheme \cite{Mazzola,Dorner} and has been experimentally 
implemented in a nuclear magnetic resonance setup \cite{Batalhao}. 
These proposals have also been extended to the open system 
scenario \cite{CampisiNJP,GooldModi}. Other proposals to 
measure work in the quantum domain relies on counting 
phonon excitations in trapped ions \cite{Huber2008,An} 
or counting electrons in single electron boxes \cite{Saira}.

Recently, two of us proposed a different method to measure work
which is based on the fact that, for quantum systems, work 
can always be measured by performing a POVM at a single 
time~\cite{Roncaglia2014}. This simple observation, that remained unnoticed
until recently, implies that work can be measured with 
a single projective measurement on an extended system. 
Thus, it is always possible to devise a measurement apparatus
that yields the work value $W$ which is a random variable 
distributed with the work probability $P(W)$.  In this paper we
will generalize this method and show how to use it to measure
work and heat in gases of cold atoms. 

There has been a lot of interest in applying 
ideas of nonequilibrium 
quantum thermodynamics in the case of isolated 
quantum many-body 
systems~{\cite{Silva, Dorosz, Dorner2012, Heyl, Sotiriadis, Haque, Fusco,Mascarenhas,Sindona}}. 
Despite the experimental advances in the field of ultracold atoms, 
an ideal platform for the quantum simulation of many-body 
systems~\cite{review_optical_lattices}, an experimentally 
feasible proposal for measuring heat and work in these 
systems is still missing. The Ramsey scheme mentioned 
earlier is based on the global coupling of an auxiliary 
two-level system with the system under consideration 
and might not be well suited for a cold atomic system. 

The proposal we present to measure work and heat in 
quantum gases generalizes the method proposed in 
\cite{Roncaglia2014} and consists in coupling the 
atoms with a continuous degree of freedom which 
can be realized by the light quadratures. The 
interaction will be chosen in such a 
way to induce a phase-space translation of the continuous 
variable position that is conditional on the value of the 
energy of the system under consideration. In short, the 
method consists of three steps: First we let atoms  
interact with light in such a way that correlations 
between them are established. Second, 
while light is stored in a quantum memory, we drive 
the atoms with the thermodynamic process we 
are interested in. Third, we retrieve the light beam 
from the memory and redirect it into the atomic ensemble
enforcing a second interaction between them. 
After these three steps, a standard homodyne detection of the output 
light is performed. The key of the method is that the statistical 
distribution of work and heat on the atoms if fully 
encoded in the statistical distribution of  the light 
quadratures. 

The paper is organized as follows: In section \ref{sec:measwork}
we present the key ingredients of the method, which generalizes
the one presented in Ref. \cite{Roncaglia2014}. Then, in the 
following sections we 
showcase two cold atoms settings where 
our proposal can be implemented using a quantum non 
demolition measurement based on the Faraday effect 
\cite{Hammerer}. The first one is designed for measuring 
work in cold atomic ensembles and is described in 
Secs.~\ref{sec:workatomic} and \ref{sec:ensemble}; 
the second example is for ultracold atoms in optical 
lattices, described in Sec.~\ref{sec:dissipated} where 
we show how to measure heat and work for the atoms. 
In the latter case, the measurement scheme allows us 
to discern if the open system dynamics is unital or not, 
by checking whether the Jarzynski equality is fulfilled. 
Finally in Sec.~\ref{sec:conclusion} we summarize. 

\section{Measuring work with a POVM}
\label{sec:measwork}

{
Let us consider a process where a quantum system with an initial state $\rho$  
is driven from an initial Hamiltonian $H$ to a final one $\tilde H$. The  
work value $W$ in each realization is defined as the energy difference 
$W=\tilde E_m - E_n$, where
$E_n$ are the eigenvalues of $H$ (i.e. $H \ket{\phi_n} =E_n\ket{\phi_n}$) and those of 
$\tilde H$ are denoted with $\tilde E_m$ (i.e., 
$\tilde H \sket{\tilde\phi_m} =\tilde E_m\sket{\tilde\phi_m}$). Thus, $W$ is a random variable 
distributed according to the following probability distribution:
$$
P(W)=\sum_{m,n}p_n p_{m|n} \,\delta[W-(\tilde E_m-E_n)],
$$
where $p_n= \bra{\phi_n}\rho\ket{\phi_n}$ and 
$p_{n|m}=|\sbra{\tilde\phi_m} U_E \ket{\phi_n}|^2$ and $U_E$ is the unitary
operation that represents the driving. As we mention in the introduction, there are many protocols that were proposed to 
experimentally reconstruct this probability distribution.} 

\begin{figure}[t!]
\begin{center}
\includegraphics[width=0.8\columnwidth]{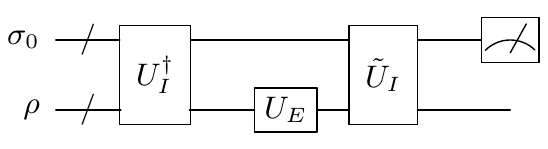}
\caption{
Quantum circuit that describes the method to measure work
as a POVM.
}
\label{fig:circuit}
\end{center}
\end{figure}

Recently an alternative method that allows to sample 
the work probability distribution has been put forward 
in ~\cite{Roncaglia2014}.
The method is based on the idea that work measurement 
is actually a POVM. As it is well known, any such generalized
measurement can be implemented as a standard projective
measurement on an enlarged system. A simple example 
of such strategy to implement the work POVM
is depicted in Fig. \ref{fig:circuit}. We assume that a 
system $\sys$ is coupled to an auxiliary system $\app$ 
in such a way that $\app$ gets entangled with $\sys$
keeping a coherent record of the energy at two 
times. In the simplest case (which will be generalized below), 
the interaction between $\sys$ and $\app$ is such that it can 
be described by the unitary evolution operators  
{$U_{I} = e^{-i \kappa P H}$} and 
$\tilde U_{I} = e^{- i  \kappa P \tilde H}$ .
The 
auxiliary system $\app$ is a continuous degree of freedom 
and $P$ is the generator of translations in the position quadrature.
In between the two entangling operations
the system is driven with the operator $U_E$. At the 
end, the ancillary system is measured in the $X$ basis and the 
moments of the $X$ variable can be estimated. The key of the 
method, as shown below, is that the distribution of results 
$P(X)$ is a coarse-grained  version of 
the full probability distribution of work.

To see how this method works we consider an 
initial thermal state for $\sys$, i.e.  
$\rho_\beta  = e^{-\beta H}/Z_\beta$ with $Z_\beta=\Tr\, e^{-\beta H}$ the partition function. In turn, for 
$\app$ we consider a general state $\sigma_0$ 
(this is a generalization of the treatment
presented in \cite{Roncaglia2014}, where $\app$
was assumed to be a position eigenstate).
The total state of the combined $\sys$--$\app$ universe
can be obtained after the sequence of evolutions
$U_{T}=\tilde U_I U_E U^\dagger_I$. 
{
Now let us see the state after each step of the algorithm. Initially the state is $\sigma_0\otimes \rho_\beta$ and an entangling operation
is applied, after this step the state can be written as
$$
U^\dagger_I(\sigma_0\otimes\rho_\beta)U_I = \sum_{i} \frac{e^{-\beta E_i}}{Z_\beta} e^{i\kappa  P E_i}\sigma_0 e^{-i\kappa  P E_i}
\otimes \sket{ \phi_i}  \sbra{\phi_i}.
$$
Then, $U_E$ is applied to system $\sys$ and, after the last entangling operation,} the final state 
$\rho_I= U_T( \sigma_0\otimes\rho_\beta)
U_T^\dagger$ can be written as
\beq
\rho_I=  
\sum_{i,m,n}
\frac{e^{-\beta E_i}}{Z_\beta} 
\chi_{m,i}\, \chi^*_{n,i}\ T_{m,i}
\sigma_0 T^\dagger_{n,i}\otimes 
{\sket{\tilde \phi_m}  \sbra{\tilde \phi_n}}, \nonumber
\eeq
where the transition elements are 
$\chi_{m,i}=\sbra{\tilde \phi_m}U_E
\ket{\phi_i}$. Moreover, the operators $T_{m,i}$
translate the state of $\app$ by an amount 
 that depends on the energy difference
$(\tilde E_m-E_i)$. Thus, they are defined as 
\beq
T_{m,i}=\exp\left[-i \kappa P (\tilde E_m \ - E_i)\right].
\nonumber
\eeq
From the total state $\rho_I$ we can compute 
the reduced density matrix of the auxiliary 
system $\rho_\app \equiv \Tr_\sys\rho_I$. Thus, 
\beq
\rho_\app=\sum_{i,m} \frac{e^{-\beta E_i}}{Z_\beta} 
|\chi_{m,i}|^2 \;\; T_{m,i}\, \sigma_0 \, T_{m,i}^\dagger.
\eeq
From this expression we can compute the 
moments of the position variable of $\app$, 
defined as $\av{X^n}{0}=\Tr[X^n \rho_I]$. They
turn out to be
\beqa
\mean{X^n}&=& 
\sum_{i,m} \frac{e^{-\beta E_i}}{Z_\beta} 
|\chi_{m,i}|^2 \Tr\left\{\sigma_0
\left[X+  \kappa({\tilde E_m-E_i})\right]^n\right\}.\nonumber  
\\
&=& 
\sum_{k=0}^{n} \binom{n}{k} \kappa^{n-k} \,
\Tr(\sigma_0 X^k) \mean{W^{n-k}}.
\eeqa
This equation establishes a simple relation between
the moments of the position variable of $\app$ and 
those of the work distribution, which are 
defined as 
$\mean{W^n}=\sum_{i,l} \frac{e^{-\beta E_i}}{Z_\beta} 
|\chi_{l,i}|^2 (\tilde E_l - E_i)^n$. 
In particular, for the first two moments the equations
are particularly simple. They read
\beqa 
\mean{X}&=&    \av{X}{0} +  \kappa \mean{W},     
\label{eq:moment1} \\
\mean{X^2} &=&  \av{X^2}{0}  +2   \kappa \av{X}{0} 
\mean{W}  + \kappa^2 \mean{W^2}.
\label{eq:moment2}
\eeqa
where $\langle\rangle_0$ denotes average on the initial state $\sigma_0$.
These equations can be used to obtain simple relations
between the dispersion (defined as 
$\Delta X^2=\av{(X-\av{X}{})^2}{}$)
and the skewness (defined as 
$\Delta X^3=\av{(X-\av{X}{})^3}{}$) of the $X$ coordinate, 
and those of the work distribution. Thus, 
\beqa
\Delta X^2&=&  \Delta X^2_{0}+ \kappa^2  \Delta W^2,\nonumber\\
\Delta X^3&=&  \Delta X^3_{0}+ \kappa^3  \Delta W^3.
\label{eq:dispersion}
\eeqa
{This also shows that the scheme can also be used to test linear response results 
which relate the dissipated energy to the variance of the work distribution \cite{Jarzynski}.}
The above equations are worth analyzing: It is clear that
the choice of the initial state $\sigma_0$ imposes strong 
constraints on the accuracy of the estimation of the properties
of the work distribution. In fact, it is clear that in order to estimate 
$\Delta W^n$ by measuring $\Delta X^n$, it is better to choose 
initial states with small dispersions. The only states for which such
dispersions vanish are the position eigenstates, which were 
considered in \cite{Roncaglia2014}. However, for a continuous
variable system such as the one we are considering here, these
states are unphysical. Instead, in this paper we will consider 
realistic scenarios for which the initial state is, typically, a 
coherent state (or a squeezed one). If instead of pure states
we use mixed ones, it is obvious that we lose accuracy. 
In fact, if the initial state is thermal (for a 
harmonic oscillator with frequency $\omega$) we have 
$\mean{ X^2}\propto \coth\left[\omega\beta/2\right]$. Therefore, 
the precision of the estimate of work 
dispersion decreases with the temperature (or, equivalently, 
to achieve the same precision in the estimate of the work
dispersion, we would need to measure the dispersion in $X$
with much higher accuracy). 

There is another generalization of the method 
presented in Ref.~\cite{Roncaglia2014} that turns out to be
useful for our purpose here. In fact, we will 
consider a more general 
interaction Hamiltonians between $\sys$ and $\app$. As 
we will show, if the Hamiltonian is non--linear in the momentum 
of $\app$ then the estimate of the moments of the work 
distribution may be simpler, and even more precise. 
To see this we consider an interaction Hamiltonian which 
induces an evolution operators given as
$U_I=e^{i  \kappa P^\alpha H}$, for integer values of
$\alpha$. In this case, it is simple to extend the previous results
and to obtain an analytic expression for the moments of the
work distribution. In fact, we find that 
\beq
\Delta X^n=  \Delta X^n_{0}+ 
\kappa^n  \Delta (W P^{\alpha-1}_{0})^n,
\eeq
a formula which is valid for $n=1,2,3$. A particularly 
simple case is attained for $\alpha=2$. 
Then, the second moment satisfy
\beq
\mean{X^2} =  
\av{X^2}{0}  + \kappa^2 \mean{W^2} \av{P^2}{0}.  \nonumber
\eeq
where we assumed $\langle X\rangle_0 =0$ as is the case of a thermal symmetric state.
This has an obvious interpretation: By considering an 
initial state which is squeezed in position we  reduce 
$\Delta X_0$. Then, the estimate of $\Delta X$ (for fixed
accuracy in the measurement of $\Delta X$) is 
higher than in the linear case. 
Again, all these results are independent of the 
initial state of the apparatus and will be useful in what follows.

\section{Work on an atomic ensemble}
\label{sec:workatomic}
In this Section we start by explaining a scheme to reconstruct the probability distribution of the work done on or extracted from a cold atomic ensemble.
The state of the ensemble, composed by $N$ 2-level atoms, can be described in terms of the collective angular momentum $\vec J$ that is the sum of the atomic spins. The components of the angular momentum operator fulfil the usual commutation relations (assuming throughout the paper that $\hbar=1$): $[J_x,J_y] = i J_z$ and all the cyclic permutations. The ensemble is subject, as in previous experiments, to a magnetic field $\vec B(t)$ that can be continuously changed in time along any direction. The Hamiltonian governing the dynamics of the ensemble is therefore:
\begin{equation}
\label{eq:Ht}
H(t)=-\gamma \vec B(t) \cdot \vec J
\end{equation}
where $\gamma$ is the gyromagnetic ratio and $\vec B(t)\equiv |\vec B|(n_x(t),n_y(t),n_z(t))$ thus we are assuming that only the direction $\vec n(t)$ of the magnetic field and not its magnitude changes in time.
The instantaneous eigenstates of $H(t)$ coincide with those of the projection of $\vec J$ along the magnetic field direction $\vec n(t)$ and we label them as $\ket{m_{\vec n(t)}}$ with eigenvalue $E_m(t) = -\gamma |\vec B|m_{\vec n(t)}$.

We now compute the work done on the atomic ensemble, initially in the state $\rho(0)$, due to the variation of the magnetic field from $\vec B (0)$ to $\vec B(\tau)$ in a time $\tau$. The ensemble state at any time can be calculated as $\rho(t)=U(t)\rho(0)U^\dagger(t)$ where we have defined the unitary evolution operator which fulfills Schr\"odinger equation:
\begin{equation}
\label{eq:sch}
i\frac{\partial}{\partial t} U(t) = H(t)U(t)
\end{equation}
with the initial condition $U(0)=\idm$. We would like to stress here that we are not making any assumption on the time variation, slow or fast, of the direction of the magnetic field.
\begin{figure}[t!]
\begin{center}
\includegraphics[width=0.8\columnwidth]{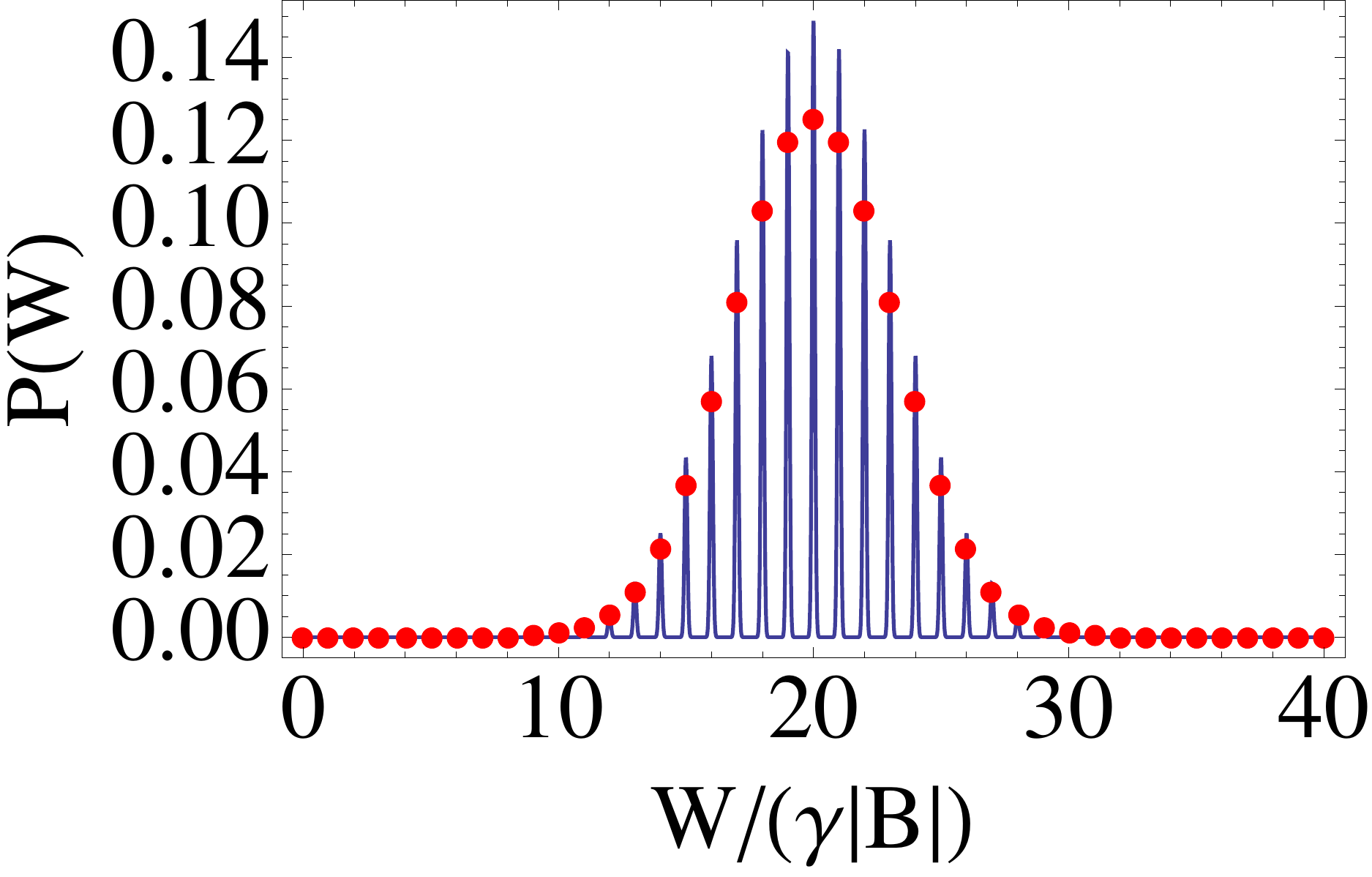}
\includegraphics[width=0.8\columnwidth]{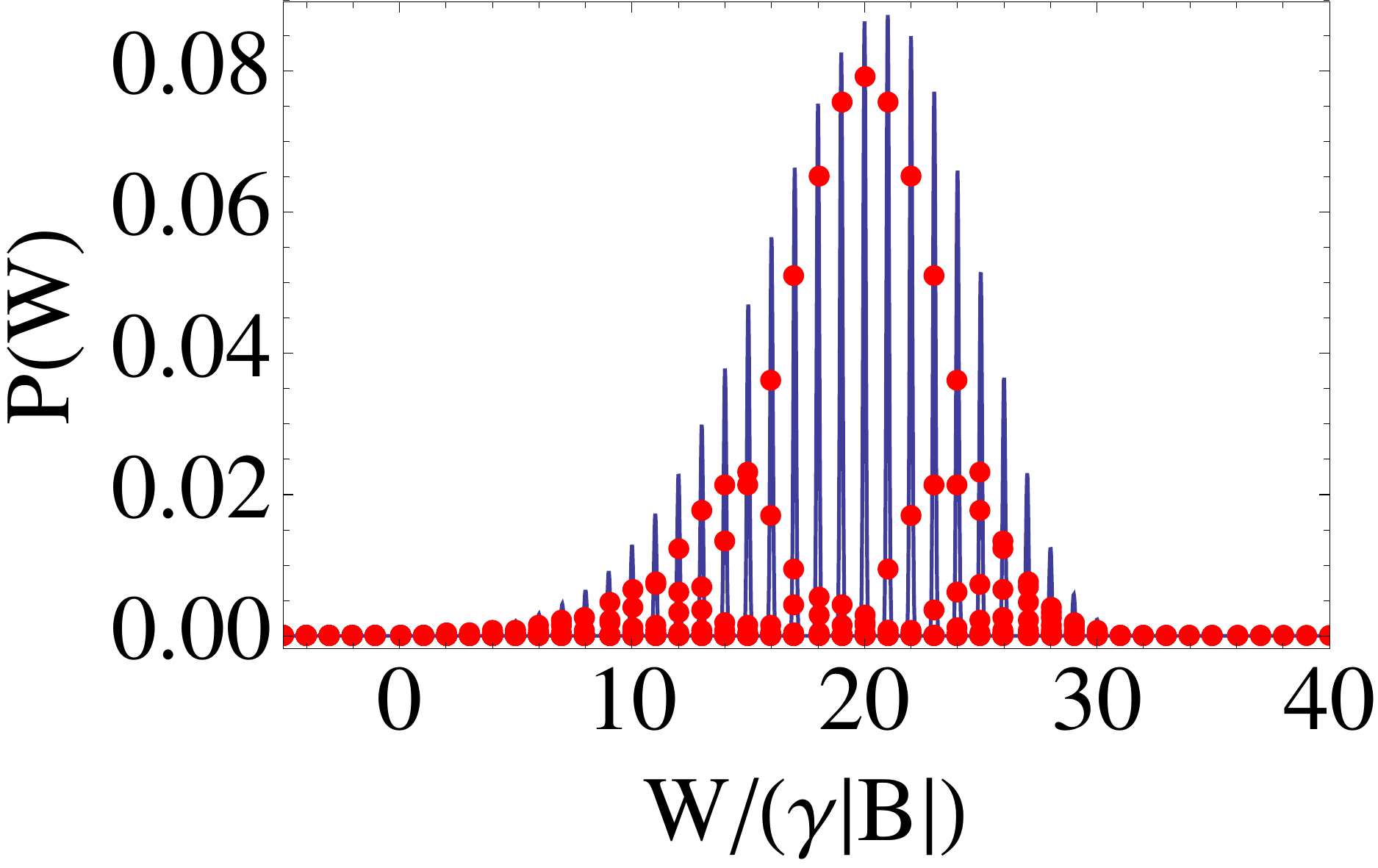}
\caption{
Work probability distribution for an atomic ensemble, with $N=40$, initially in thermal equilibrium  with a magnetic field pointing along the $z$ direction and instantaneously rotated to the $y$ axis. We consider an ensemble initially at zero temperature ($\beta=+\infty$, top) and one with inverse temperature $\beta=1$ (bottom). Dots represent the strength of the Dirac delta function from definition \eqref{eq:PW}. The blue solid line is an appropriately rescaled continuous coarse grained version of the WPD. 
}
\label{fig:exactW}
\end{center}
\end{figure}

Taking as a definition the two time protocol,  work $W$ is a classical stochastic variable with probability distribution:
\begin{equation}
\label{eq:PW}
P(W) = \sum_{m,m'} p_m p_{m'|m} \delta\left[W-E_{m'}(\tau) + E_m(0)\right]
\end{equation}
where $p_m = \bra {m_{\vec n(0)}} \rho(0) \ket{m_{\vec n(0)}}$ is the probability to find the initial state in the initial Hamiltonian eigenstate $\ket{m_{\vec n(0)}}$ and 
$$p_{m'|m} = \left|\; \left\langle {m'_{\vec n(\tau)}} \left | U(\tau) \right |{m_{\vec n(0)}} \right\rangle \; \right|^2$$
 is the conditional probability that evolving with the evolution operator $U(\tau)$ the initial Hamiltonian eigenstate $\ket{m_{\vec n(0)}}$ the state of the system is found, at time $\tau$, in the final Hamiltonian eigenstate $\ket{m'_{\vec n(\tau)}}$.

We start with a simple case where we assume that the initial magnetic field is pointing along the $z$ direction and, at $t=0$, is instantaneously rotated to the $y$ axis, thus $U(\tau)=\idm$. We assume that the ensemble is initially in thermal equilibrium with inverse temperature $\beta$ (assuming the Boltzmann constant $k_B=1$) so that its state is:
\begin{equation}
\rho(0) = \frac {1}{Z(0)} e^{-\beta H(0)}
\end{equation}
 where $Z(0) = \Tr\, e^{-\beta H(0)}$ is the initial partition function ensuring the normalisation of the state density matrix. 
In this case the work probability distribution (WPD) depends on the overlaps $ | \langle {m'_y}|{m_z} \rangle  |^2$ between the angular momentum eigenstates along the $z$ and $y$ directions. These can be calculated in terms of the Wigner D-matrix but the result is cumbersome and will not be reported here. The results for the WPD can be found in Fig.~\ref{fig:exactW}. 
It can be observed that for very low temperatures the probability distribution resembles a Gaussian function. This can be explained as follows. The initial state is polarised along the $z$ direction, so each spin is in a superposition of the up and down states along the $y$ axis. As the total state is the tensor product of each spin wavefunction, the resulting distribution is binomial, thus approaching a Gaussian shape for large number of atoms. More precisely, for a large number of particles, and using Holstein-Primakoff approximation, the atomic state can be regarded as a coherent state. For the instantaneous quench we are considering, the WPD depends only on the transition probabilities $p_{m'|m}$ which, in the Holstein-Primakoff picture, represents the wave function squared of such coherent state, therefore a Gaussian function, its position-like operator being proportional to the angular momentum $J_y$ along the final magnetic field.

For large temperatures this is not true anymore, and other transitions from initial excited states acquire a higher weight. These give rise to many more peaks distorting the WPD to a skewed function. We have studied the normalised skewness $\rm{Skew}[W]$ of the WPD as a function of temperature. The normalised skewness is defined as:
\begin{equation}
\rm{Skew}[W] = \left\langle \left(\frac{W-\langle W\rangle}{\sigma_W}\right)^3  \right\rangle
\end{equation}
where $\sigma_W$ is the work standard deviation.

\begin{figure}[t]
\begin{center}
\includegraphics[width=0.8\columnwidth]{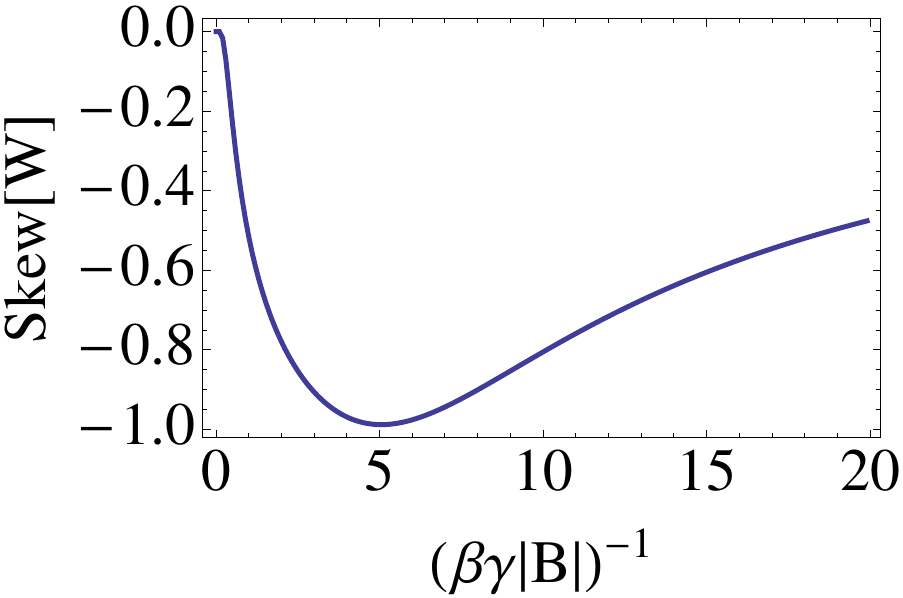}
\caption{
Skewness of the probability distribution of the work done on an atomic ensemble with $N=40$ atoms after an instantaneous rotation of its magnetic field from the direction $z$ to the direction $y$. 
}
\label{fig:skew}
\end{center}
\end{figure}

The results, reported in Fig.~\ref{fig:skew}, show that the skewness is always negative meaning that, although most of the probability is located to the right of the maximum of the distribution, there is a long tail of small probabilities to the left of the maximum. This is not uncommon for the WPD \cite{Fusco, Huber2008} and sometimes it gives rise to non-zero probability for negative work values. Fig.~\ref{fig:skew} shows also an interesting result: the skewness approaches zero for very small, as we said earlier, or very large temperatures. 
In the large temperature limit the skewness also approaches zero because the initial state is proportional to the identity meaning that all energy eigenstates are equally probable. This leads to a symmetric distribution as the transitions probabilities are symmetric: $|\langle{m_z}\ket{m_y}|^2=|\langle{-m_z}\ket{m_y}|^2=|\langle{m_z}\ket{-m_y}|^2=|\langle{-m_z}\ket{-m_y}|^2$. Thus $P(W)$ is symmetric around zero and the skewness reduces to zero.
For intermediate temperatures $(\beta \gamma|B|)^{-1}\sim 5$ there is a maximum of the absolute value of the skewness.

We now consider a slow quench of the magnetic field and calculate the work done on the ensemble for different speeds $\omega=1/\tau$. We therefore assume that the magnetic field rotates at constant angular speed $\omega$ as:
\begin{equation}
\vec B(t) =|\vec B|( \cos\omega t \hat{\vec k} + \sin\omega t\hat{\vec \jmath} ) 
\end{equation}
For this particular choice the eigenenergies $E_m$ do not depend on time and there are no degeneracies. We thus expect that for sufficiently small angular speed $\omega$ the evolution to be (quantum) adiabatic: since there are no transitions induced by the time variation of the Hamiltonian, the state populations do not change in time and the state at all times remains in thermal equilibrium. In this regime we expect the average work $\langle W\rangle $ to approach the free energy difference $\Delta F$ which, for the process we consider, is null. For higher speed $\omega$ we expect the process to excite the system and bring it out of equilibrium. This in turn produces irreversible work defined as:
\begin{equation}
W_{\rm irr} = \langle W\rangle - \Delta F = \langle W\rangle
\end{equation}
where the last equality follows from our assumptions that the modulus of the magnetic field does not change.

\begin{figure}[t]
\begin{center}
\includegraphics[width=0.8\columnwidth]{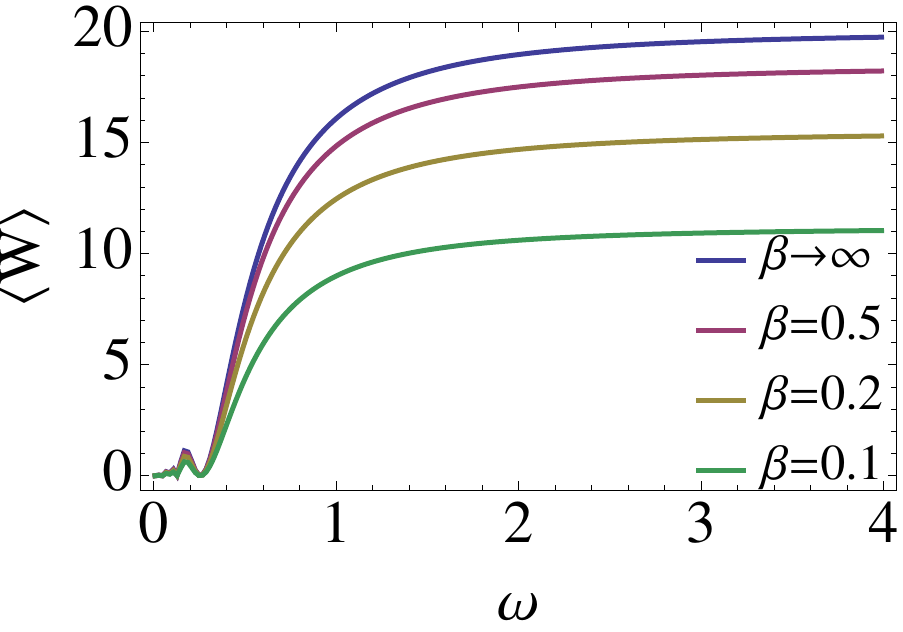}
\caption{
Average work done on the atomic ensemble for rotating the magnetic field form the $z$ to the $y$ direction at constant angular speed $\omega$ and for different temperatures $1/\beta$. As before, we used $N=40$.
}
\label{fig:W_vs_speed}
\end{center}
\end{figure}

The results for $\langle W\rangle$ are shown in Fig.~\ref{fig:W_vs_speed}. As we expected, for very small $\omega$ the average work tends to zero while growing and approaching a limiting value for very fast quenches. This value coincides with the average work calculated assuming instantaneous quenches $U(\tau)=\idm$. The figure also shows the dependence of the average work for different temperatures. For high temperatures the average work reduces as the system initially occupies many excited states. In the limit of infinite temperature, the initial state of the system is the unitary invariant completely mixed state proportional to the identity. In this limit, the average work is zero because any transformation leaves the state unaltered.

\section{Reconstructing the work distribution using light}
\label{sec:ensemble}
\subsection{The scheme}
We propose a scheme, following Ref.~\cite{Roncaglia2014}, to experimentally reconstruct the probability distribution of the work done on an atomic ensemble when varying the applied magnetic field. To this end we use a light-matter interface based on the Faraday rotation~\cite{Kupriyanov}. If light polarised along the $x$ axis propagates along the YZ plane and illuminates the atomic ensemble at an angle $\alpha$ with the $z$ axis, the interaction Hamiltonian reads:
\begin{equation}
\label{eq:Hint}
H_I (\alpha) = \frac{a}{T} S_z (\cos\alpha J_z+\sin\alpha J_y)
\end{equation}
where $a$ is the coupling constant and $T$ is the duration of the pulse.
The Stokes operators are defined as:
\begin{eqnarray}
  S_x&=&\frac{1}{2}(  a^\dagger_x   a_x-  a^\dagger_y   a_y),
  \\
  S_y&=&\frac{1}{2}(  a^\dagger_y   a_x+  a^\dagger_x   a_y),
  \\
  S_z&=&\frac{1}{2i}(  a^\dagger_y   a_x-  a^\dagger_x   a_y)
\end{eqnarray}
where the operators $a_x$ and $a_y$ annihilates a photon with polarisation along $x$ and $y$, respectively.
We assume that the light pulse is strongly polarised along the $x$ axis: $S_x\approx\langle S_x\rangle = N_{ph}/2$ where $N_{ph}$ is the number of photons. Within this approximation, we can treat the Stokes operators in the two perpendicular directions as conjugated variables: $S_z = P\sqrt{N_{ph}/2}$ and $S_y = X\sqrt{N_{ph}/2}$, so that $[X,P]=i$.

Using these assumptions the evolution operator corresponding to a pulse with Hamiltonian \eqref{eq:Hint} is:
\begin{equation}
U_I(\alpha) = \exp\left[-i \kappa P J(\alpha) \right ]
\end{equation}
where $\kappa = a\sqrt{N_{ph}/2}$ and $J(\alpha) = (\cos\alpha J_z+\sin\alpha J_y)$. With atomic ensemble at room temperatures the coefficient $\kappa$ could be very small for our purposes, as we would need a value $\kappa\approx 1$.
For ultracold atoms the optical depth, and therefore $\kappa$, could be made larger although results in this direction have not yet been demonstrated.
We could also write the transformation $U_I(\alpha)$ as:
\begin{equation}
U_I(\alpha) = \exp\left[i \tilde\kappa P H_A(\alpha) \right ]
\end{equation}
where $H_A(\alpha) = \gamma|\vec B| J(\alpha)$, which is equivalent to $H(t)$ in Eq.~\eqref{eq:Ht}, and we set $\tilde\kappa=\kappa/(\gamma|\vec B|)$.
Thus it is clear that transformation $U_I(\alpha)$ is a spatial translation of the continuous state of light conditional on 
the atomic ensemble energy. It is this conditional interaction that makes it possible to read the WPD from the state of the light.

\begin{figure}[t!]
\begin{center}
\includegraphics[width=0.8\columnwidth]{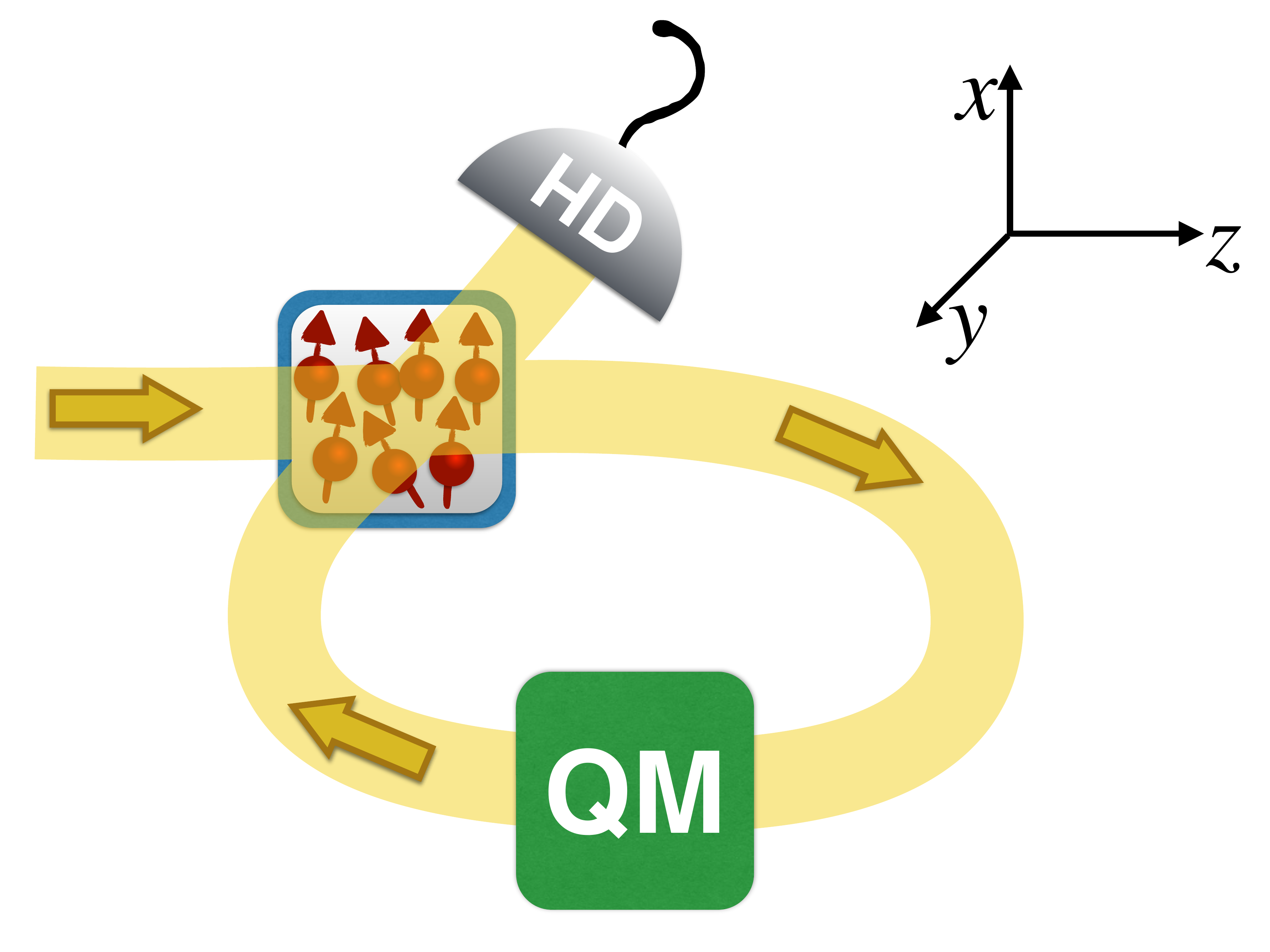}
\caption{
Proposed setup to measure the probability distribution of the work done on an atomic ensemble. A beam of light strongly polarised along the $x$ axis propagates along the $z$ direction illuminating the atomic ensemble thus reading the initial energy. The beam is then stored in a quantum memory (QM) while the magnetic field of the ensemble is changed in time. Finally the beam is retrieved from the quantum memory and let pass through the ensemble along the negative $y$ direction. The polarisation fluctuations of the emerging beam are then measured using homodyne detection (HD).
}
\label{fig:ensemble-setup}
\end{center}
\end{figure}

We now follow the idea from \cite{Roncaglia2014}. Initially the polarisation fluctuation state of the light is assumed to be characterised by a Gaussian wave function, $\ket{X=0}_L$ centred in zero with variance $\sigma^2$. Although the vacuum would correspond to $\sigma^2=1/2$ we carry on our analysis for generic $\sigma$ thus encompassing also squeezed states. For the sake of simplicity we assume that the atoms are initially in a pure state $\ket\psi_A$, but the same exact scheme works also for mixed states.

As illustrated in Fig.~\ref{fig:ensemble-setup}, the protocol consists in shining the atoms with a laser beam, strongly polarised along $x$ and propagating along a direction on the $yz$ plane and forming an angle $\alpha_0+\pi$ with the $z$ axis. During this first step, 
light and atoms interact with a Hamiltonian proportional to $H_A(\alpha_0)$.
While the beam is stored in a quantum memory, the atoms undergo the process during which the magnetic field is rotated eventually pointing to the direction in the $yz$ plane forming an angle $\alpha_1$ with the $z$ axis. The atomic state is evolved with evolution operator $U(t)$ fulfilling Eq.~\eqref{eq:sch}. Finally, the light beam is retrieved from the quantum memory and let pass through the atoms along a direction forming an angle $\alpha_1$ with the $z$ axis. During this step light and atoms  interact with a Hamiltonian proportional to $H_A(\alpha_1)$.
Thus, at the end the state of the light encodes the difference between the final and initial energy for each posible quantum trajectory.
It is at the very end, when the measurement is performed, that coherence is destroyed. In this way the method samples $W$ with probability
$P(W)$.

Mathematically the state of atoms and light before light is measured is
\begin{eqnarray}
\ket \Psi_{AL} &=& U_I(\alpha_1) U(t) U_I^\dagger(\alpha_0) \ket{\psi}_A\ket{X=0}_L =
\\ 
&=&\sum_{m,m'} c_m c_{m'|m}\;\; \ket{E_{m'}}_A\ket{X=\kappa(E_{m'}-E_m)}_L
\nonumber
\end{eqnarray}
where $c_m = {}_A\bk{E_m}{\psi}_A$ and $c_{m'|m} =  {}_A\bra{E_{m'}} U(t) \ket{E_m}_A$ and 
where states like $\ket{X=\kappa(E_{m'}-E_m)}_L$ represent the initial state of light rigidly translated by the quantity $\kappa(E_{m'}-E_m)$.

The reconstructed work distribution can be found from the probability density distribution of the $X$ quadrature of light (assuming no degeneracies):
\begin{equation}
\label{eq:PL}
P_L(X) = \sum_{m,m'} p_m p_{m'|m} \frac{1}{\sqrt{2\pi\sigma^2}}\exp\left[-\frac{(X-\kappa(E_{m'}-E_m))^2}{2\sigma^2}  \right]
\end{equation}
where we have identified $p_m=|c_m|^2$ and $p_{m'|m}=|c_{m'|m}|^2$. Notice the difference with the work distribution in Eq.~\eqref{eq:PW}: apart from the conversion factor $\kappa$ the light distribution corresponds to a coarse grained version of $P(W)$ where Dirac delta functions have been replaced by Gaussians with width $\sigma$. We therefore expect a faithful reconstruction of the WPD when $\sigma/\kappa$ is sufficiently smaller than the energy change $E_{m'}-E_m$.

Using Eqs. \eqref{eq:moment1} and \eqref{eq:moment2} we can obtain the first two moments of the light distribution:
\beq
\label{eq:mean}
\langle X\rangle = \kappa\langle W\rangle
\eeq
And for the second moment:
\begin{eqnarray}
\langle X^2\rangle = \sigma^2+ \kappa^2\langle W^2\rangle
\end{eqnarray}
so that the variance of the light distribution is:
\begin{equation}
\label{eq:variance}
\Delta X^2 = \sigma^2 +\kappa^2\Delta W^2
\end{equation}

Therefore provided that $\kappa$ is sufficiently strong we can estimate the first two moments of the work distribution by measuring the light fluctuations.
A similar two- or multiple-passage protocol has been previously discussed in Refs.~\cite{Sherson2006} for the implementation of a quantum memory.

\subsection{An example}

To showcase our proposal we consider the process described in Sec.~\ref{sec:workatomic}. The atoms are initially in thermal equilibrium and subject to a magnetic field along the $z$ direction. The magnetic field is suddenly rotated to the $y$ direction and we want to reconstruct the WPD of this process and compare it with the exact one calculated in Sec.~\ref{sec:workatomic}.

\begin{figure}[t!]
\begin{center}
\includegraphics[width=0.8\columnwidth]{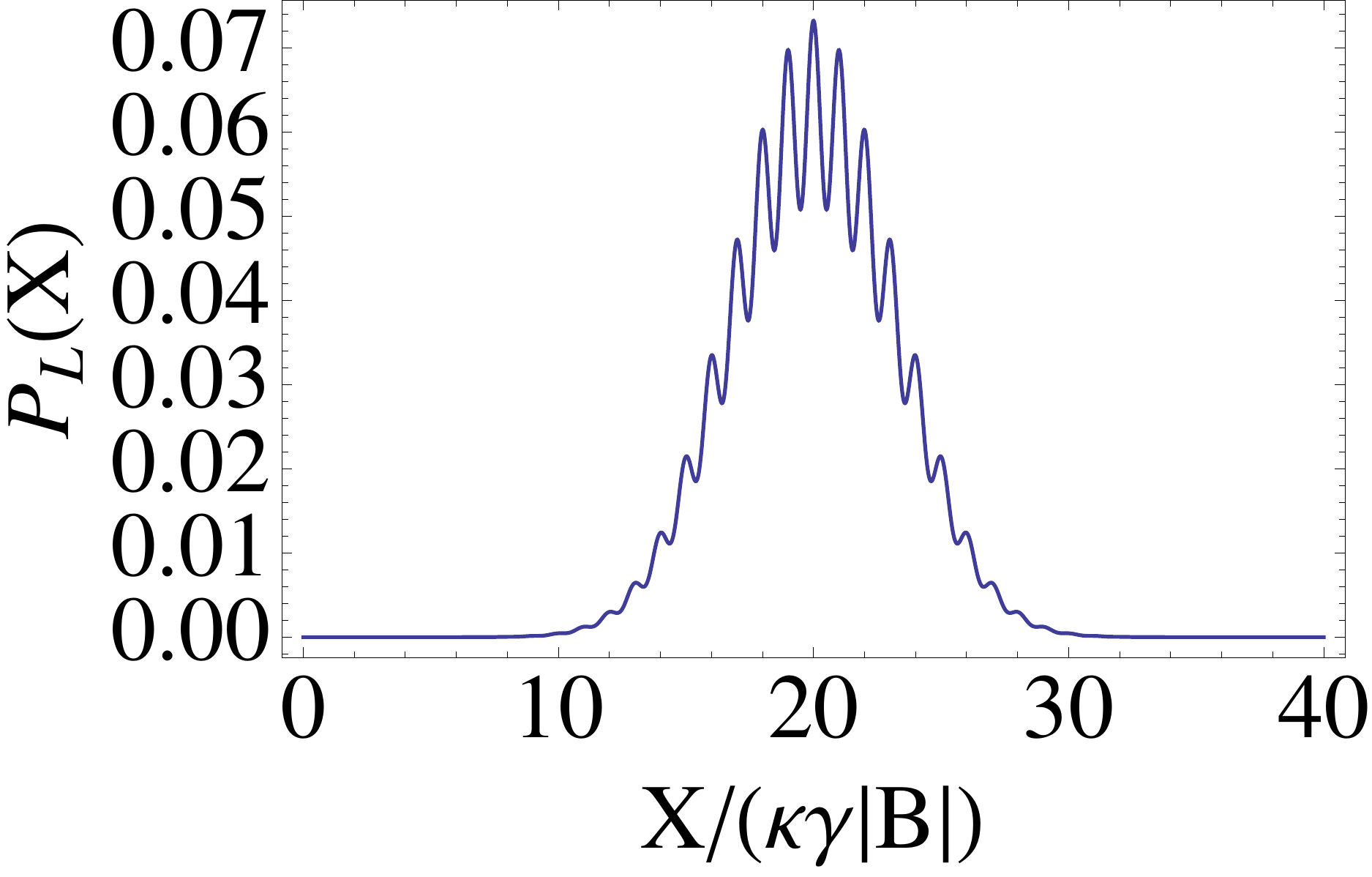}
\includegraphics[width=0.8\columnwidth]{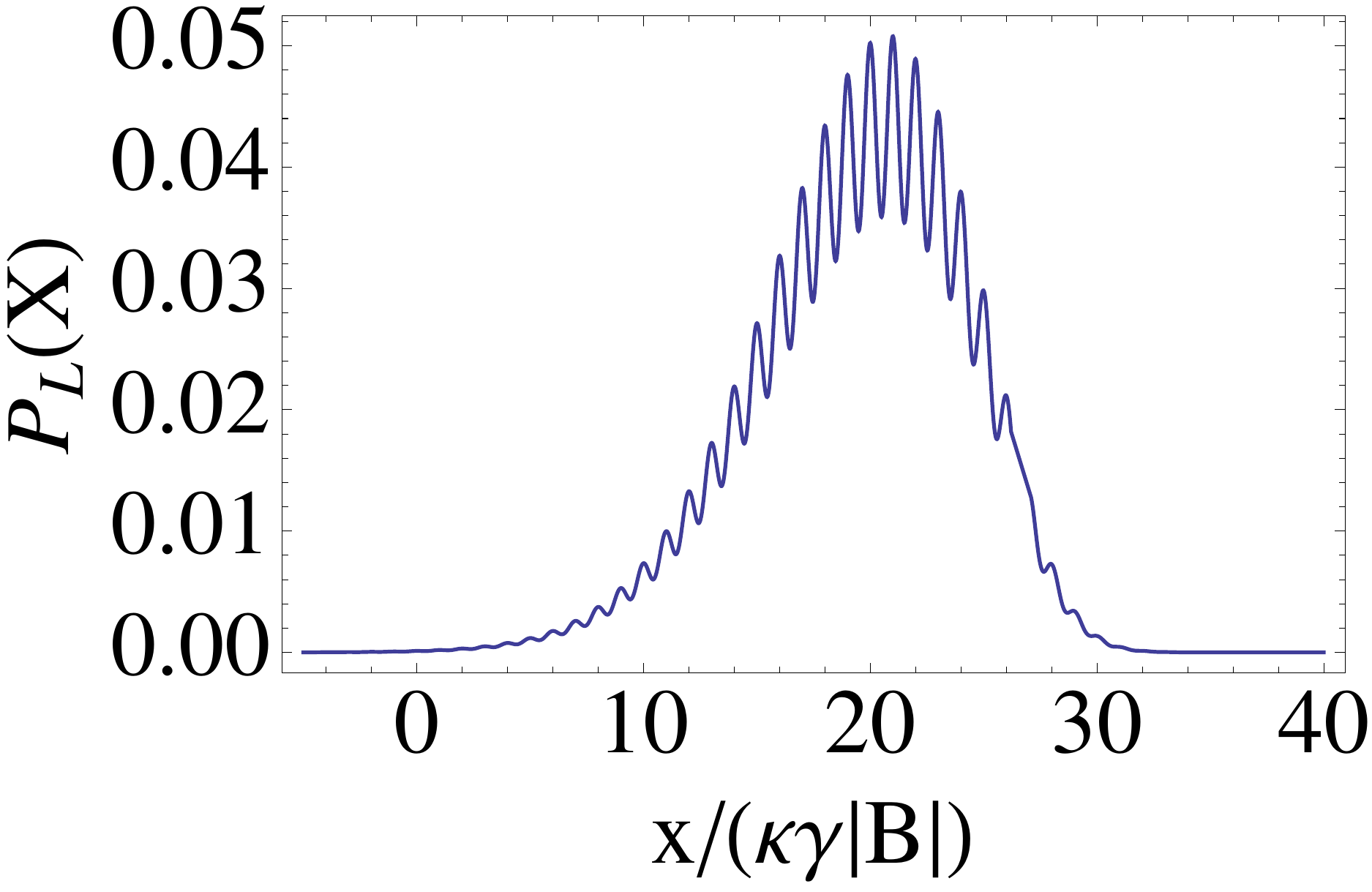}
\caption{Light quadrature probability distribution for an atomic ensemble initially polarised along the negative $z$ direction and instantaneously quenched along the $y$ axis. The initial temperature of the ensemble is zero (Top) and $\beta=1$ (Bottom). Parameters: $N=40; \kappa=2; \sigma^2=1/2$.}
\label{fig:prob}
\end{center}
\end{figure}

The distribution for the light quadrature $X$ can be found by inserting in Eq.~\eqref{eq:PL} the expressions for $p_m$ and $p_{m'|m}$ used in Sec.~\ref{sec:workatomic}.
The light probability distribution is shown in Fig.~\ref{fig:prob} for zero temperature and for a temperature $\beta^{-1}=1$. The light distribution is the sum of narrow Gaussians at each of the red points of Fig.~\ref{fig:exactW}. Therefore it represents a coarse grained version of it. Nevertheless, all the important features such as the first few moments and the overall shape agree with the exact result.
So even with modest resources like using coherent states $\sigma^2=1/2$ and a coupling $\kappa=2$ it is possible to reconstruct quite faithfully the work probability distribution.

The ultimate test of our reconstructed WPD is Jarzynski equality \cite{Jarzynski}
\begin{equation}
\label{eq:jar}
\langle e^{-\beta W} \rangle = e^{-\beta \Delta F}=1
\end{equation}
where the last equality follows from the fact that for us the free energy difference $\Delta F$ is zero.
  
Since the work variable $W$ corresponds to the renormalised quadrature $X/\kappa$ we compute:
\begin{eqnarray}
\langle e^{-\beta  X/\kappa} \rangle &=& \int_{-\infty}^\infty e^{-\beta X/\kappa} P_L(X) dX=
\\ \nonumber
&=& \sum_{m,m'} p_m p_{m'|m} e^{-\beta(E_{m'}-E_m)} 
\exp\left[\frac{\sigma^2\beta^2}{2\kappa^2}\right ] =
\\ \nonumber 
&=& \langle e^{-\beta W} \rangle \exp\left[\frac{\sigma^2\beta^2}{2\kappa^2}\right ] = \exp\left[\frac{\sigma^2\beta^2}{2\kappa^2}\right ]
\end{eqnarray}
where in the last equality we used Jarzynski relation \eqref{eq:jar}.
Thus, Jarskynski equality is estimated with a correction that decreases with the coupling $\kappa$ and the temperature and decreases with the width $\sigma$ of the initial light polarisation state. 
A similar results was found for generalised energy measurements \cite{Watanabe}.

\begin{figure}[t]
\begin{center}
\includegraphics[width=0.8\columnwidth]{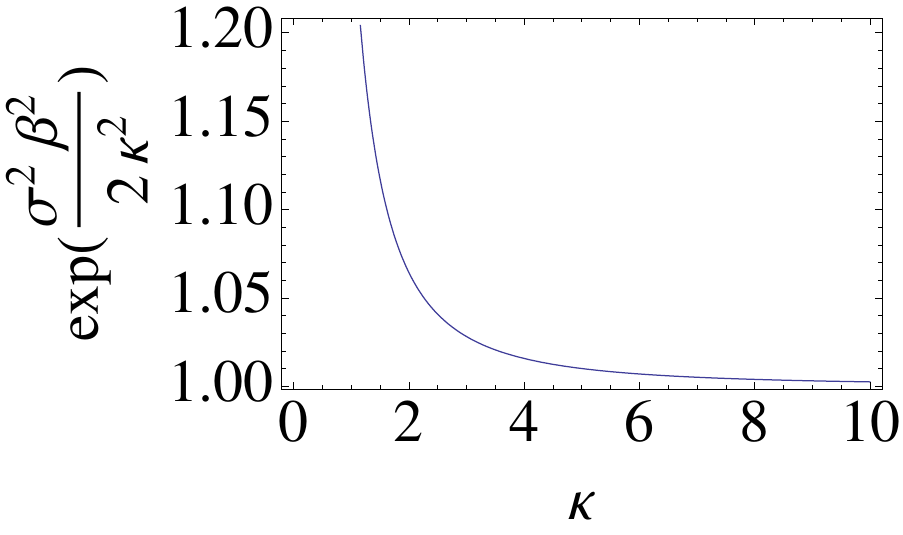}
\caption{Correcting factor to the Jarzynski equality as a function of the coupling constant $\kappa$. Parameters: $N=40; \sigma^2=1/2; \beta=1$.}
\label{fig:jarcorrection}
\end{center}
\end{figure}

 Using the following parameters: $N=40; \kappa=2; \sigma^2=1/2; \beta=1$, we obtain
 \begin{equation}
\langle e^{-\beta X/\kappa} \rangle = 1.06
\end{equation}
which is only 6$\%$ from the expected result.
A plot of the correcting factor $\exp\left[\frac{\sigma^2\beta^2}{2\kappa^2}\right ]$ is shown in Fig.~\ref{fig:jarcorrection} where it is clear that values of $\kappa$ above 5 gives a negligible correction to the Jarzynski equality.

\section{Measuring dissipated energy in an open system}
\label{sec:dissipated}

\subsection{Generalities on fluctuation relations in open quantum system}
So far we have discussed a method to reconstruct the probability distribution of work done or extracted from an isolated system. We now extend the method to a non unitary evolution in which the system $S$ is coupled to an environment $E$ during the process. Relaxing the assumptions of unitary processes, we have to be careful when talking about work. The system in fact exchanges energy, which we may well call heat, with the environment. Thus it is more accurate to talk about the energy change $\mathcal E$ of the system and its fluctuation relations \cite{Goold_Modi}. We are assuming as in the previous section that we perform a two-time energy measurement on the system, the only difference is that now the system evolution is not unitary.

In the open quantum system scenario, it is common to introduce a complete positive trace preserving (CPTP) map $\Phi$ acting on the initial density matrix $\rho_S(0)$ (assuming no initial system-environment correlations) to obtain the evolved density matrix at time $t$. If the evolution of the combined system plus environment is unitary and governed by the operator $U_{SE}$ the map can be expressed as:
\begin{equation}
\rho_S(t) =\Phi[\rho_S(0)]={\rm Tr}_E\left[ U_{SE} \rho_S(0)\otimes\rho_E(0) U_{SE}^\dagger  \right]
\end{equation}
The map can be conveniently cast in terms of Kraus operators:
\begin{equation}
\Phi[\rho_S(0)] = \sum_k A_k \rho_S(0) A_k^\dagger 
\end{equation}
where, due to the trace preserving nature of the map, the Kraus operators fulfil $ \sum_k A_k^\dagger A_k  = \idm$. It is possible to define a dual map $\Phi^*$ as
\begin{equation}
\Phi^*[\rho_S(0)] = \sum_k A^\dagger_k \rho_S(0) A_k
\end{equation}
which however is not in general trace preserving. A map $\Phi$ is called unital if the corresponding dual map $\Phi^*$ is trace preserving. This condition is equivalent to requiring that $\Phi$ maps the completely mixed state $\idm_S$ into itself: $\Phi(\idm_S) = \idm_S$.

It has been shown before \cite{Rastegin,Zanardi} that when calculating an analogous relation to Jarzynki's one obtains a result that depends on the dual map:
\begin{equation}
\langle e^{-\beta \mathcal E} \rangle = {\rm Tr}\left[ \rho_S(0) \Phi(\idm) \right]
\end{equation}
where $\rho_S(0) =\frac{ e^{-\beta H_S}}{\Tr\, e^{-\beta H_S}}$ and the quantity on the right-hand side has been called efficacy of the process. Thus, Jarzynski equality for the energy change is fulfilled, i.e. the right hand side is 1 as we are considering zero free energy change, if and only if the map $\Phi$ is unital.

\subsection{An example with atomic spins in optical lattices}
\begin{figure}[t]
\begin{center}
\includegraphics[width=0.95\columnwidth]{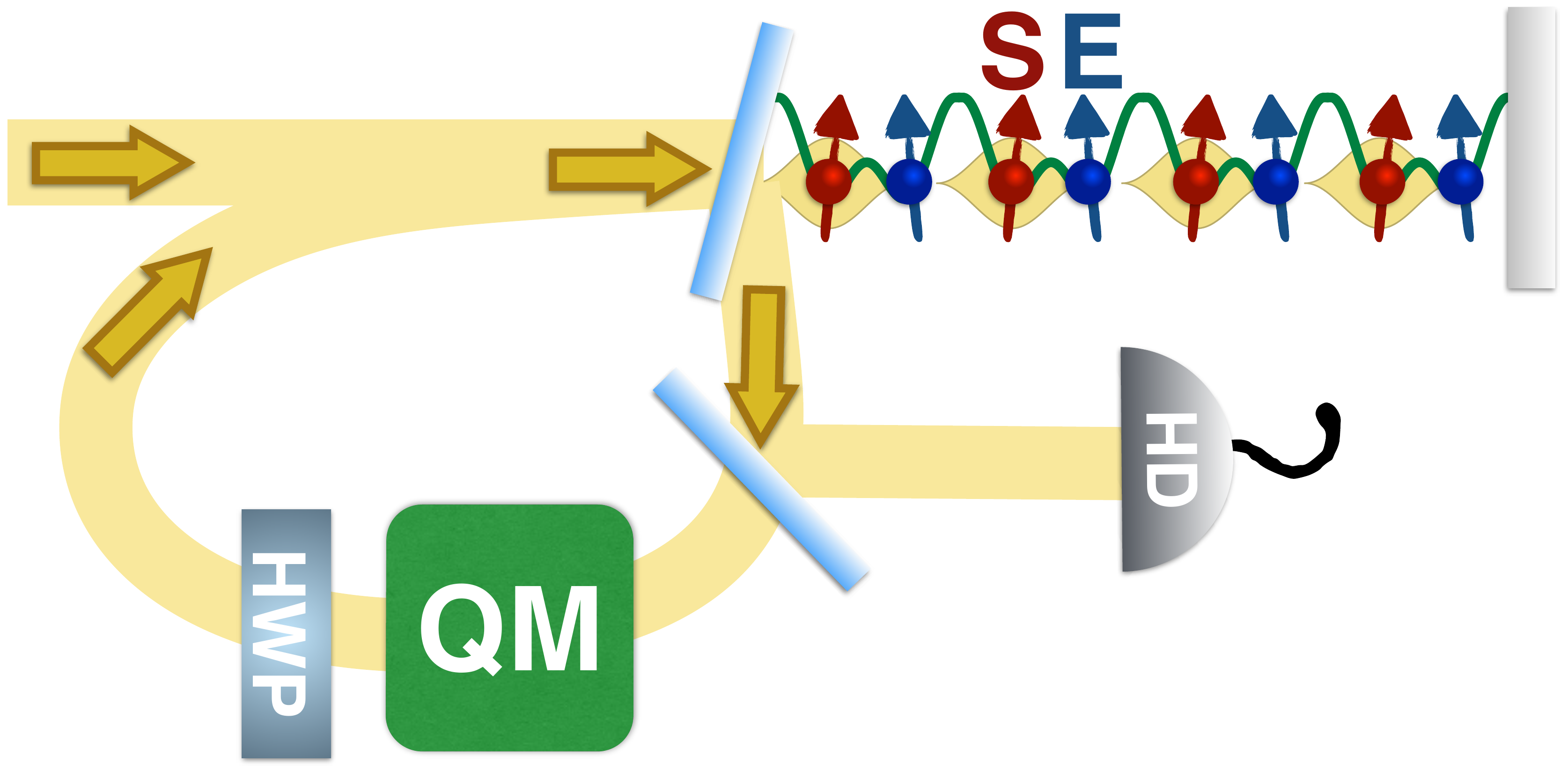}
\caption{Scheme for measuring energy dissipated in an array of spins trapped in an optical lattice. The lattice is formed by an array of double wells where a single atomic spin occupies each of the two wells. We consider the spin in the left well as the system and the spin in the right as the environment. A laser pulse (yellow) is first shone onto the atoms in a standing wave configuration created by a mirror so that it illuminates only the system atoms. The light pulse is then stored in a quantum memory (QM) until a unitary transformation between system and environment spins is generated. Then the beam is retrieved, passes through a half-wave plate where its polarisation is rotated by 180 degrees and is redirected to the atoms again thus completing the reading protocol. Finally, the laser pulse is analysed with homodyne detection (HD).}
\label{fig:setupspins}
\end{center}
\end{figure}

To test the ideas discussed in the previous paragraphs, we consider the setup sketched in Fig.~\ref{fig:setupspins}. A superlattice potential of double wells is created with the aid of two standing waves with wave vectors having a ratio of 2. For large enough intensities, and assuming no vacancies, each well will contain exactly one atom, i.e. the system is in a Mott insulator with unit filling. Probing ultracold atoms in superlattice potentials has been proposed in \cite{Ben}. We assume the atom sitting in the left well to be the system and the atom in the right well to be the environment. In this limit tunnelling is suppressed and a super-exchange interaction between the pseudo-spin internal levels can be induced by lowering the barrier between the two wells.
The spins are initially in thermal equilibrium at the same temperature:
\begin{equation}
\rho_R(0) =\frac{ e^{-\beta H_R}}{{\rm Tr} e^{-\beta H_R}}, \quad R=S,E
\end{equation}
with $H_R=B\sigma^R_x$ and $S,E$ indicates the system and environment spins, respectively. 
 
To measure energy change $\mathcal E$, we first measure the initial energy of the system by projecting the initial density matrix on the eigenstates $\ket +$ and $\ket -$  of $H_S$. Then the thermodynamic process consists in coupling system and environment with the XXZ interaction:
\begin{equation}
\label{eq:HSE}
H_{SE} = \sigma^S_x\sigma^E_x +\sigma^S_y\sigma^E_y+\Delta\sigma^S_z\sigma^E_z
\end{equation}
and evolving it in time for a time $t$ with the evolution operator $U_{SE}=\exp[-iH_{SE}t]$. In Eq.~\eqref{eq:HSE}, $\Delta$ is the interaction anisotropy which can be tuned by accurately changing the atoms scattering length near a Feshbach resonance~\cite{review_optical_lattices}. 

We then consider the reduced density matrix of the system only and measure again the energy $H_S$. The probability distribution of the energy change $\mathcal E$ is:
\begin{eqnarray}
\label{eq:probE}
P(\mathcal E) &=& \bra +\rho_S(0)\ket +\;  \bra +\rho_+(t)\ket + \delta(\mathcal E)
\\
&+&\bra +\rho_S(0)\ket + \; \bra -\rho_+(t)\ket - \delta(\mathcal E-2B)
\nonumber
\\
&+&\bra -\rho_S(0)\ket - \; \bra +\rho_-(t)\ket + \delta(\mathcal E+2B)
\nonumber
\\
&+&\bra -\rho_S(0)\ket - \; \bra -\rho_-(t)\ket - \delta(\mathcal E)
\nonumber
\end{eqnarray}
where 
\begin{equation}
\rho_\pm(t) = {\rm Tr}_E\left[ e^{-i H_{SE} t} \ket\pm\bra\pm\otimes\rho_E(0) e^{i H_{SE} t}  \right]
\end{equation}

It is easy to check that the probability distribution Eq.~\eqref{eq:probE} is normalised. From Eq.~\eqref{eq:probE} is it possible to write an analogous of Jarzinsky equality:
\begin{equation}
\label{eq:expE}
\langle e^{-\beta \mathcal E} \rangle = 1+\tanh^2(\beta B)\sin(2t)\sin(2\Delta t)
\end{equation}
Thus, not surprisingly we get a time-dependent correction to Jarzinsky equality due to the openness of the system evolution.

However, notice that for $\Delta=0$, corresponding to the XXZ model, the Jarzynski equality is fulfilled at all times. This is because the CPTP map $\Phi$ that evolves the system becomes unital for  $\Delta=0$. 
In fact, applying the map to the completely mixed state, we obtain:
\begin{equation}
\Phi\left[\frac {\idm_S}{2}\right]=\frac 12
 \left(\begin{array}{cc}
 1 & g(t)
 \\
g(t) &  1
 \end{array}\right)
\end{equation}
where $g(t)=\tanh(\beta B)\sin(2t)\sin(2\Delta t)$.
Thus the 1-norm of the difference of $\Phi\left[ {\idm_S}/{2}\right]$ from the identity is related to the violation of the Jarzynski equality:
\begin{equation}
\left\|\Phi\left[\frac {\idm_S}{2}\right]-\frac {\idm_S}{2}\right\|_1 = \frac 12|g(t)|.
\end{equation}

\subsection{Scheme to reconstruct energy change in optical lattices}
We finally consider the reconstruction of the dissipated energy distribution with the Faraday rotation scheme. 
As shown in Fig.~\ref{fig:setupspins} the scheme is very similar to the one with an atomic ensemble. This time, the pulse produces a standing wave with a double period with respect to the optical lattice. This means that only the left-most spin in each double well is strongly illuminated by the light probe. In this way we can measure the total energy of all the identical system spins in the lattice.

The pulse is first sent through the atomic array and then stored in a quantum memory, as before. In this first stage the light polarisation fluctuation contains information about the initial energy of the system. Then, while the light is stored, the atoms interact according to the XXZ interaction described before. After this, the light pulse is retrieved from the memory, its polarisation is rotated by 180 degrees by a half-wave plate, and let pass through the atoms again, thus reading the final energy of the system spins. The pulse is finally analysed with a homodyne detection. 

As in Sec.~\ref{sec:ensemble}, the reconstructed distribution is obtained from Eq.\eqref{eq:probE} with the substitution:
\begin{equation}
\delta(\mathcal E) \to \frac{1}{\sqrt{2\pi}\sigma}\exp\left[-\frac{1}{2\sigma^2} (X-\kappa\mathcal E)^2\right]
\end{equation}
Thus the probability distribution of the light quadrature $X$ is a coarse-grained version of the true energy change probability distribution.
The average exponentiated energy for a single spin is corrected by a factor:
\begin{eqnarray}
\label{eq:expElight}
\langle e^{-\beta X/\kappa} \rangle =e^{\frac{\sigma^2\beta^2}{2\kappa^2}} \langle e^{-\beta \mathcal E} \rangle 
\end{eqnarray}
therefore if $\sigma\ll\kappa$ the reconstruction is possible. Notice that when the map is unital ($\Delta=0$) the reconstructed Jarzynski quantity is time independent. Therefore even if the reconstructed result differs from the correct one, from its time dependence, it unambiguously signals the unitality of the map.

So far, we have calculated the dissipated energy distribution for a pair of spins. As the light interacts with all the $N$ pairs of atoms in the lattice, the total dissipated energy is the sum of all the energies of each system atom. As these behave independently the joint probability distribution is factorised, so that the expectation value of the exponential becomes the $N$th power of the results in \eqref{eq:expE} and \eqref{eq:expElight}.

\section{Conclusions}
\label{sec:conclusion}
 In summary, we have proposed an experimentally feasible method to reconstruct the full distribution of the energy change, specifically work and heat, of ultracold atomic gases. Although our proposal employs a light-matter interface based on the quantum Faraday rotation, we stress that it could be adapted to other similar setups, for example a Bose-Einstein condensate in a cavity.
Finally, our proposal is able to reveal fundamental properties of non-unitary evolutions that can be exploited for quantum environment engineering.

\acknowledgments
We thank S. Deffner, J. Goold, J. Sherson for fruitful discussions. GDC acknowledges hospitality at Universidad de Buenos Aires where part of this work was carried on.
GDC acknowledges support by the UK EPSRC (EP/L005026/1), the John Templeton Foundation (grant ID 43467), the EU Collaborative Project
TherMiQ (Grant Agreement 618074) and the COST Action MP1209. AJR and JPP acknowledge support from ANPCyT (PICT-2010-02483, PICT-2013-0621), CONICET and UBACyT.



\begin{thebibliography}{99}
\bibitem{unital} {A unital map on a $C^{*}$-algebra is a map $\phi$ which preserves the identity element:
$\phi ( \idm ) = \idm$. }
\end{thebibliography}
\end{document}